\documentclass[prc,preprint,showpacs,preprintnumbers,amsmath,amssymb]{revtex4}

\usepackage[mathscr]{eucal}
\usepackage{graphicx}
\def\slr#1{\setbox0=\hbox{$#1$}           
   \dimen0=\wd0                                 
   \setbox1=\hbox{/} \dimen1=\wd1               
   \ifdim\dimen0>\dimen1                        
      \rlap{\hbox to \dimen0{\hfil/\hfil}}      
      #1                                        
   \else                                        
      \rlap{\hbox to \dimen1{\hfil$#1$\hfil}}   
      /                                         
   \fi}

\def\ksq{k^2}

\def\mytint#1{\!\int\!\!\frac{d^3\!{#1}}{(2\pi)^3}\,}
\def\gev#1{ GeV${}^{#1}$}
\def\be{\begin{eqnarray}}
\def\ee{\end{eqnarray}}

\renewcommand{\theequation}%
    {\arabic{section}.\arabic{equation}}
\makeatletter \@addtoreset{equation}{section} \makeatother

\begin{document}

\preprint{BCCNT: 04/01/321}

\title{Chiral Quark Model Calculation of the Momentum Dependence of Hadronic
Current Correlation Functions at Finite Temperature}

\author{Xiangdong Li}
\affiliation{%
Department of Computer System Technology\\
New York City College of Technology of the City University of New
York\\
Brooklyn, New York 11201 }%

\author{Hu Li}
\author{C. M. Shakin}
\email[email:]{casbc@cunyvm.cuny.edu}
\author{Qing Sun}
\author{Huangsheng Wang}

\affiliation{%
Department of Physics and Center for Nuclear Theory\\
Brooklyn College of the City University of New York\\
Brooklyn, New York 11210
}%

\date{January, 2004}

\begin{abstract}
We calculate spectral functions associated with hadronic current
correlation functions for vector currents at finite temperature.
We make use of a model with chiral symmetry, temperature-dependent
coupling constants and temperature-dependent momentum cutoff
parameters. Our model has two parameters which are used to fix the
magnitude and position of the large peak seen in the spectral
functions. In our earlier work, good fits were obtained for the
spectral functions that were extracted from lattice data by means
of the maximum entropy method (MEM). In the present work we extend
our calculations and provide values for the three-momentum
dependence of the vector correlation function at $T=1.5T_c$. These
results are used to obtain the correlation function in coordinate
space, which is usually parametrized in terms of a screening mass.
Our results for the three-momentum dependence of the spectral
functions are similar to those found in a recent lattice QCD
calculation for charmonium [S. Datta, F. Karsch, P. Petreczky and
I. Wetzorke, hep-lat/0312037]. However, we do not find the
expontential behavior in coordinate space that is usually assumed
for the spectral function for $T>T_c$ and which allows for the
definition of a screening mass.
\end{abstract}


\maketitle

\section{INTRODUCTION}
In a number of recent works [\,1-\,3] we have calculated various
hadronic correlation functions and compared our results to results
obtained in lattice simulations of QCD [\,4-\,6]. The lattice
results for the correlators, $G(\tau, T)$, may be used to obtain
the corresponding spectral functions, $\sigma(\omega, T)$, by
making use of the relation \be G(\tau, T)=\int_0^\infty d \omega
\sigma(\omega, T) K(\tau, \omega, T)\,,\ee where \be K(\tau,
\omega, T)=\frac{\cosh[\omega(\tau-1/2T)]}{\sinh(\omega/2T)}\,.\ee
The procedure to obtain $\sigma(\omega, T)$ from the knowledge of
$G(\tau, T)$ makes use of the maximum entropy method (MEM)
[\,7-9\,], since $G(\tau, T)$ is only known at a limited number of
points.

In our studies of meson spectra at $T=0$ and at $T<T_c$ we have
made use of the Nambu--Jona-Lasinio (NJL) model. The Lagrangian of
the generalized NJL model we have used in our studies is

\begin{flushleft}
\be \mathcal L&=&\overline{q}(i\slr\gamma-m^0)q+\frac{\overline
G_S}{2}\sum_{i=0}^8 [(\overline{q} \lambda^{i} q)^2+(\overline{q}
i \gamma_5 \lambda^{i} q)^2]\\\nonumber &-&\frac{\overline
G_V}{2}\sum_{i=0}^{8}[(\overline{q} \lambda^{i}\gamma_\mu
q)^2+(\overline{q} \lambda^{i}\gamma_5\gamma_\mu q)^2]\\\nonumber
&+&\frac{G_D}{2} \lbrace
\det[\overline{q}(1+\lambda_5)q]+\det[\overline{q}(1-\lambda_5)q]\rbrace
+\mathcal L_{conf}\,. \ee
\end{flushleft}

Here, $m^0$ is a current quark mass matrix, $m^0=diag(m_u^0,
m_d^0, m_s^0)$. The $\lambda_i$ are the Gell-Mann (flavor)
matrices and $\lambda^0=\sqrt{2/3}\mathbf{1}$, with $\mathbf{1}$
being the unit matrix. The fourth term is the 't Hooft interaction
and $\mathcal L_{conf}$ represents the model of confinement used
in our studies of meson properties.

In the study of hadronic current correlators it is important to
use a model which respects chiral symmetry, when $m^0=0$.
Therefore, we make use of the Lagrangian of Eq. (1.3), while
neglecting the 't Hooft interaction and $\mathcal L_{conf}$. In
order to make contact with the results of lattice simulations we
use the model with the number of flavors, $N_f=1$. Therefore, the
$\lambda^i$ matrices in Eq. (1.3) may be replaced by unity. We
then use \be \mathcal
L&=&\overline{q}(i\slr\gamma-m^0)q+\frac{G_S}{2}[(\overline{q}q)^2+(\overline{q}
i \gamma_5 q)^2]\\\nonumber &-&\frac{G_V}{2}[(\overline{q}
\gamma_\mu q)^2+(\overline{q}\gamma_5\gamma_\mu q)^2] \ee in order
to calculate the hadronic current correlation functions. Thus,
there are essentially three parameters to consider, $G_S$, $G_V$
and a Gaussian cutoff parameter $\alpha$, which restricts the
momentum integrals through a factor
$\exp[-\overrightarrow{k}^2/\alpha^2]$. When we use the NJL model
to study matter at finite temperature, we introduce the
temperature-dependent parameters $G_S(T)$, $G_V(T)$ and
$\alpha(T)$. These parameters have been adjusted to obtain fits to
the spectral functions $\sigma(\omega, T)$ for $T/T_c=1.5$ and
$T/T_c=3.0$, which are the values of $T/T_c$ studied in the
lattice simulations of QCD [\,10\,].

The temperature-dependent coupling constants and cutoff parameters
of our work are analogous to the corresponding density-dependent
parameters introduced in Ref. [11] and [12]. Further study of
models with temperature-dependent and density-dependent parameters
are of interest and a general theoretical formalism for the
introduction of such dependencies should be considered.

In this work we limit our study to the data for the vector
correlator at $T=1.5T_c$ [10, 13] and therefore only need to
specify $G_V(T)$ and $\alpha(T)$ at that temperature. In Figs. 1
and 2 we show the data obtained by the MEM method at $T/T_c=1.5$
and $T/T_c=3.0$ for both pseudoscalar and vector correlators [10,
13]. The second peaks in these correlators are known to be a
lattice artifacts [13] and our calculations do not reproduce those
peaks.

The organization of our work is as follows. In Appendix A we
present the formalism for calculation of pseudoscalar and vector
correlation functions making use of the Lagrangian of Eq. (1.4).
In Appendix A we discuss the calculation of the correlator in the
case the quark and antiquark carry zero total momentum. In
Appendix B we show how the formalism is modified for the
correlator calculated at finite momentum $\overrightarrow P$. In
Section II we present the results of our calculation of the
imaginary part of the correlator $\sigma(\omega, \overrightarrow
P)$. (Since we place $\overrightarrow P$ along the \emph{z}-axis
this quantity may be written as $\sigma(\omega, 0, 0, P_z)$ in
accord with the notation of Ref. [10].) Our results for
$\sigma(\omega, \overrightarrow P)$ will be presented for a series
of values of $|\overrightarrow P|$ in Section II. In Section III
we present our result for the coordinate-dependent correlator
$C(z)$ which is proportional to the correlator defined in Eq. (1)
of Ref. [10], \be
C(z)=\frac12\int_{-\infty}^\infty\,dP_ze^{iP_zz}\int_0^\infty\,d\omega\frac{\sigma(\omega,
0, 0, P_z)}\omega\,. \ee We may also use the form \be
C(z)=\frac14\int_{-\infty}^\infty\,dP_ze^{iP_zz}\int_0^{P^2}\,dP^2\,\frac{\sigma(P^2,
0, 0, P_z)}{P^2}\,. \ee Finally, in Section IV we present our
conclusion and further discussion.

\section{temperature dependent hadronic current correlators at finite
momentum}

We make use of the formalism presented in Appendices A and B to
obtain values of the vector correlator at $T=1.5T_c$. Here we take
$T_c=270$ MeV, since we are considering lattice calculations made
in the quenched approximation.

In Fig. 3 we present $\sigma(\omega)/\omega^2$, for various values
of $|\overrightarrow P|$, as function of $\omega/T$. Comparison
may be made to the lattice data shown in Fig. 2 [10, 13]. (We
again note that our calculation does not reproduce the second peak
in the lattice data which is known to be a lattice artifact [13].)
The curves shown in Fig. 3 are given for values of
$|\overrightarrow P|$ ranging from 0.10 GeV to 3.10 GeV in steps
of 0.20 GeV. The same data are presented in Fig. 4, where our
results for $\sigma(\omega)/\omega^2$ are given as a function of
$\omega^2$. In this figure we see that the maximum value of
$\omega^2$ used in our calculations is 90\gev2, so that
$\omega_{max}=9.49$ GeV. Our results for the various values of
$|\overrightarrow P|$ given in Fig. 3 may be compared to Fig. 20
of Ref. [14]. We see that the results calculated by completely
different methods are similar.

\begin{figure}
\includegraphics[bb=0 0 280 235, angle=0, scale=1]{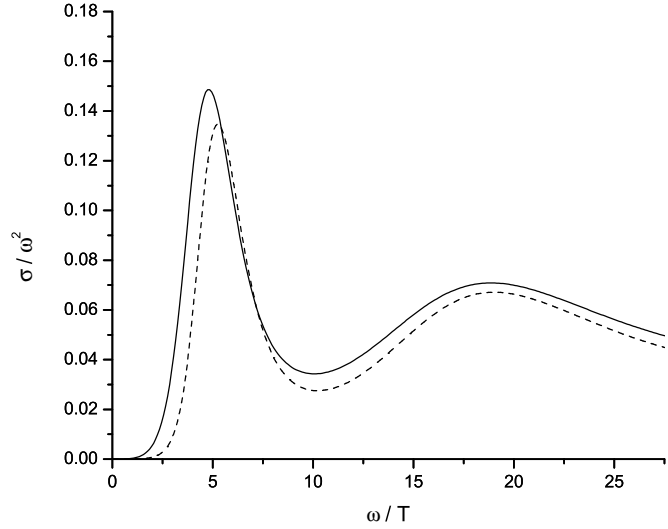}%
\caption{The spectral functions $\sigma/\omega^2$ for pseudoscalar states
obtained by
MEM are shown [\,10, 13\,]. The solid line is for $T/T_c=1.5$ and the dashed
line is
for $T/T_c=3.0$. The second peak is thought to be a lattice artifact [13]
and is not
reproduced in our model. (See Fig. 3, for example.)}
\end{figure}

\begin{figure}
\includegraphics[bb=0 0 280 235, angle=0, scale=1]{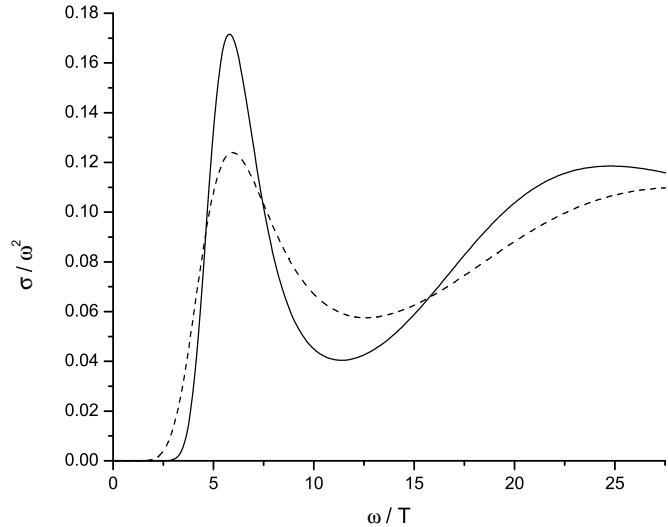}%
\caption{The spectral functions $\sigma/\omega^2$ for vector states obtained
by MEM
are shown [\,10, 13\,]. See the caption of Fig. 1. These results may be
compared to
the theoretical results shown in Fig. 3 where we see the absence of the
second peak.
That peak is thought to be a lattice artifact [13].}
\end{figure}

\begin{figure}
\includegraphics[bb=0 0 280 235, angle=0, scale=1]{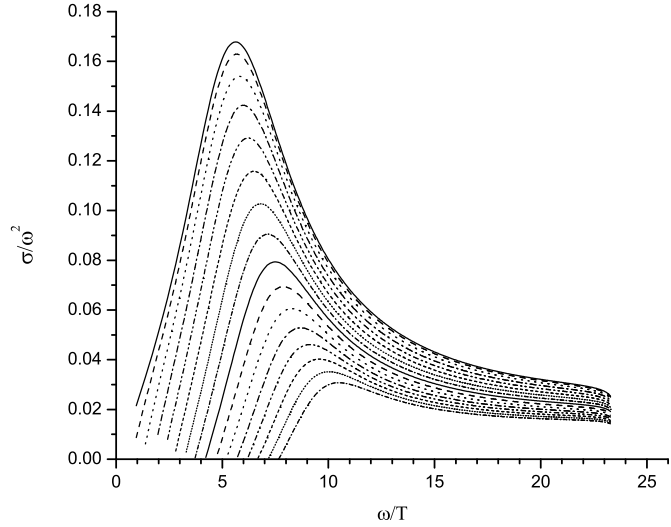}%
\caption{The imaginary part of the correlator $\sigma(\omega)/\omega^2$ is
shown for
various values of $|\vec P|$ as a function of $\omega/T$. Starting with
the topmost curve the values of $|\vec P|$ in GeV units are 0.10, 0.30,
0.50,
0.70, 0.90, 1.10, 1.30, 1.50, 1.70, 1.90, 2.10, 2.30, 2.50, 2.70, 2.90, and
3.10. Here we
have used $G_S=0.66$ \gev{-2} and $\alpha=4.4$ GeV.}
\end{figure}

\begin{figure}
\includegraphics[bb=0 0 280 235, angle=0, scale=1]{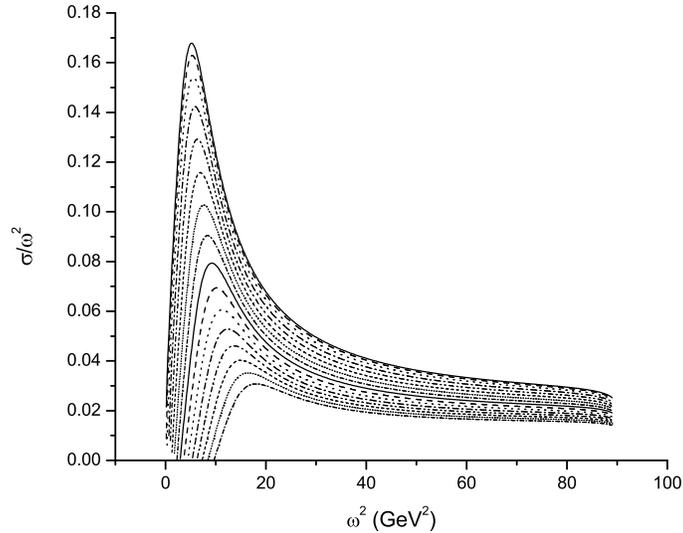}%
\caption{The imaginary part of the vector correlation function
$\sigma(\omega)/\omega^2$ is shown for various values of $|\vec P|$
as a function of $\omega^2$. See the caption of Fig. 3.}
\end{figure}

\begin{figure}
\includegraphics[bb=0 0 280 235, angle=0, scale=1]{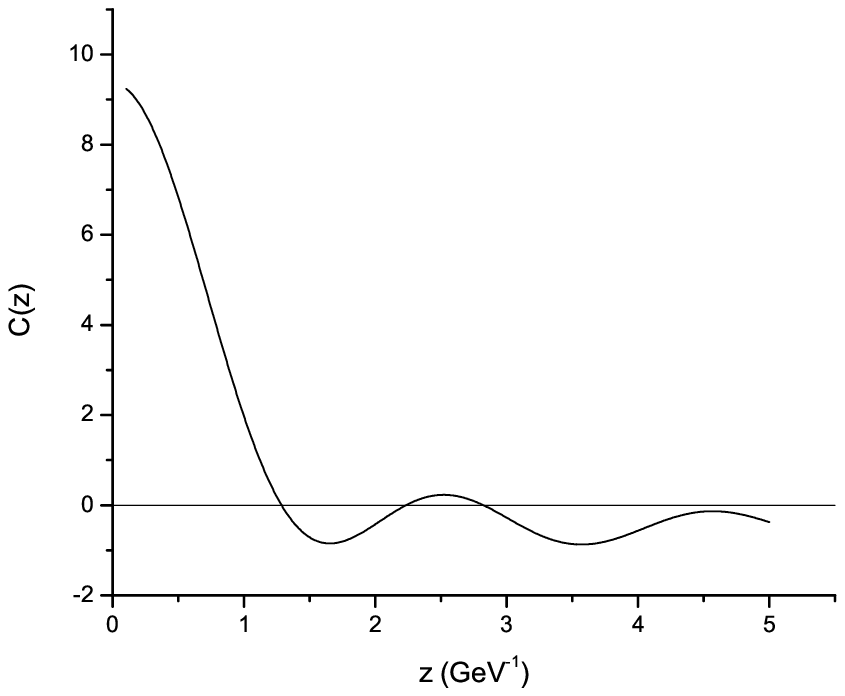}%
\caption{The correlation function $C(z)$ defined in Eq. (1.6) is
shown.}
\end{figure}

\section{the coordinate-space correlation function}

We have used the results of our calculations which were presented
in Fig. 4 to calculate $C(z)$ of Eq. (1.6). The result of that
calculation is shown in Fig. 5. We note that the simple assumption
for the behavior of this correlator that is usually made,
$C(z)\thicksim\exp[-m_{sc}z]$, is not born out in our calculation.
Therefore, we do not provide a value for the screening mass
$m_{sc}$ which is defined when assuming exponential behavior of
the correlator. For the study of charmonium on the lattice, values
found for the screening mass are given in Ref. [14].

\section{discussion}

Recent theoretical work concerning the quark-gluon plasma has been
discussed by Shuryak [15]. He notes that the physics of excited
matter produced in heavy ion collision in the region $T_c<T<4T_c$
is different from that of a weakly coupled quark-gluon plasma.
That is due to the strong coupling generated by the bound states
of the quasiparticles. These bound states appear as resonant
structures when the imaginary parts of the hadronic correlators
are extracted from lattice data using the maximum entropy method
(MEM) [4-9]. These resonances lead to very strong interactions
between the quasiparticles and are, in part, responsible for very
small mean free paths and collective flow in heavy-ion collisions.
That flow may be described by hydrodynamics. Indeed, Shuryak
suggests that ``... if the system is macroscopically large, then
its description via \emph{thermodynamics} of its bulk properties
(like matter composition) and \emph{hydrodynamics} for space-time
evolution should work" [15].

In our present work we have presented a simple chiral model that
is able to reproduce the resonances extracted from the lattice
data via the MEM procedure. We have calculated the imaginary parts
of the vector correlator for various values of the total external
momentum $\vec P$. We have also calculated the correlation
function, $C(z)$, and do not find the simple exponential behavior
$C(z)\thicksim\exp[-m_{sc}z]$ usually assumed for that correlator.
In the special case in which the correlator is dominated by a
simple delta function for small $\omega$, we may define $m_{sc}$
and see that the screening and pole masses are equal [10].)

\appendix
  \renewcommand{\theequation}{A\arabic{equation}}
  \setcounter{equation}{0}  
  \section{}  

For ease of reference, we present a discussion of our calculation
of hadronic current correlators taken from Ref.\,[\,3\,]. The
procedure we adopt is based upon the real-time finite-temperature
formalism, in which the imaginary part of the polarization
function may be calculated. Then, the real part of the function is
obtained using a dispersion relation. The result we need for this
work has been already given in the work of Kobes and Semenoff
[\,16\,]. (In Ref.\,[\,16\,] the quark momentum is $k$ and the
antiquark momentum is $k-P$. We will adopt that notation in this
section for ease of reference to the results presented in
Ref.\,[\,16\,].) With reference to Eq.\,(5.4) of Ref.\,[\,16\,],
we write the imaginary part of the scalar polarization function as
\be \mbox{Im}\,J_S(\textit{P}\,{}^2,
T)=\frac12N_c\beta_S\,\epsilon(\textit{P}\,{}^0)\mytint ke^{-\vec
k\,{}^2/\alpha^2}\left(\frac{2\pi}{2E_1(k)2E_2(k)}\right)\\\nonumber
\times\{[1-n_1(k)-n_2(k)]
\delta(\textit{P}\,{}^0-E_1(k)-E_2(k))\\\nonumber-[n_1(k)-n_2(k)]
\delta(\textit{P}\,{}^0+E_1(k)-E_2(k))\\\nonumber-[n_2(k)-n_1(k)]
\delta(\textit{P}\,{}^0-E_1(k)+E_2(k))\\\nonumber-[1-n_1(k)-n_2(k)]
\delta(\textit{P}\,{}^0+E_1(k)+E_2(k))\}\,.\ee Here,
$E_1(k)=[\,\vec k\,{}^2+m_1^2(T)\,]^{1/2}$. Relative to Eq.\,(5.4)
of Ref.\,[\,16\,], we have changed the sign, removed a factor of
$g^2$ and have included a statistical factor of $N_c$. In
addition, we have included a Gaussian regulator, $\exp[\,-\vec
k\,{}^2/\alpha^2\,]$. The value $\alpha=0.605$ GeV was used in our
applications of the NJL model in the calculation of meson
properties at $T=0$. We also note that \be
n_1(k)=\frac1{e^{\,\beta E_1(k)}+1}\,,\ee and \be
n_2(k)=\frac1{e^{\,\beta E_2(k)}+1}\,.\ee For the calculation of
the imaginary part of the polarization function, we may put
$\ksq=m_1^2(T)$ and $(k-P)^2=m_2^2(T)$, since in that calculation
the quark and antiquark are on-mass-shell. In Eq.\,(A1) the factor
$\beta_S$ arises from a trace involving Dirac matrices, such that
\be \beta_S&=&-\mbox{Tr}[\,(\slr k+m_1)(\slr k-\slr P+m_2)\,]\\
&=&2P^2-2(m_1+m_2)^2\,,\ee where $m_1$ and $m_2$ depend upon
temperature. In the frame where $\vec P=0$, and in the case
$m_1=m_2$, we have $\beta_S=2P_0^2(1-{4m^2}/{P_0^2})$. For the
scalar case, with $m_1=m_2$, we find \be \mbox{Im}\,J_S(P^2,
T)=\frac{N_cP_0^2}{8\pi}\left(1-\frac{4m^2(T)}{P_0^2}\right)^{3/2}
e^{-\vec k\,{}^2/\alpha^2}[\,1-2n_1(k)\,]\,,\ee where \be \vec
k\,{}^2=\frac{P_0^2}4-m^2(T)\,.\ee

For pseudoscalar mesons, we replace $\beta_S$ by
\be \beta_P&=&-\mbox{Tr}[\,i\gamma_5(\slr k+m_1)i\gamma_5(\slr k-\slr
P+m_2)\,]\\
&=&2P^2-2(m_1-m_2)^2\,,\ee which for $m_1=m_2$ is $\beta_P=2P_0^2$
in the frame where $\vec P=0$. We find, for the $\pi$ mesons, \be
\mbox{Im}\,J_P(P^2,T)=\frac{N_cP_0^2}{8\pi}\left(1-\frac{4m^2(T)}{P_0^2}\right)^{1/2}
e^{-\vec k\,{}^2/\alpha^2}[\,1-2n_1(k)\,]\,,\ee where $ \vec
k\,{}^2={P_0^2}/4-m_u^2(T)$, as above. Thus, we see that the phase
space factor has an exponent of 1/2 corresponding to a
\textit{s}-wave amplitude. For the scalars, the exponent of the
phase-space factor is 3/2, as seen in Eq.\,(A6).

For a study of vector mesons we consider \be
\beta_{\mu\nu}^V=\mbox{Tr}[\,\gamma_\mu(\slr k+m_1)\gamma_\nu(\slr
k-\slr P+m_2)\,]\,,\ee and calculate \be
g^{\mu\nu}\beta_{\mu\nu}^V=4[\,P^2-m_1^2-m_2^2+4m_1m_2\,]\,,\ee
which, in the equal-mass case, is equal to $4P_0^2+8m^2(T)$, when
$m_1=m_2$ and $\vec P=0$. This result will be needed when we
calculate the correlator of vector currents. Note that, for the
elevated temperatures considered in this work, $m_u(T)=m_d(T)$ is
quite small, so that $4P_0^2+8m_u^2(T)$ can be approximated by
$4P_0^2$, when we consider the vector current correlation
functions. In that case, we have \be \mbox{Im}\,J_V(P^2,T) \simeq
\frac{2}{3}\mbox{Im}\,J_P(P^2,T)\,.\ee At this point it is useful
to define functions that do not contain that Gaussian regulator:
\be\mbox{Im}\,\tilde{J}_P(P^2,T)=\frac{N_cP_0^2}{8\pi}\left(1-\frac{4m^2(T)}{P_0^2}\right)^{1/2}[\,1-2n_1(k)\,]\,,\ee
and
\be\mbox{Im}\,\tilde{J}_V(P^2,T)=\frac{2}{3}\frac{N_cP_0^2}{8\pi}\left(1-\frac{4m^2(T)}{P_0^2}\right)^{1/2}[\,1-2n_1(k)\,]\,,\ee
For the functions defined in Eq.\,(A14) and (A15) we need to use a
twice-subtracted dispersion relation to obtain
$\mbox{Re}\,\tilde{J}_P(P^2,T)$, or
$\mbox{Re}\,\tilde{J}_V(P^2,T)$. For example,
\be\mbox{Re}\,\tilde{J}_P(P^2,T)=\mbox{Re}\,\tilde{J}_P(0,T)+
\frac{P^2}{P_0^2}[\,\mbox{Re}\,\tilde{J}_P(P_0^2,T)-\mbox{Re}\,\tilde{J}_P(0,T)\,]\\\nonumber
+\frac{P^2(P^2-P_0^2)}{\pi}\int_{4m^2(T)}^{\tilde{\Lambda}^{2}}
ds\frac{\mbox{Im}\,\tilde{J}_P(s,T)}{s(P^2-s)(P_0^2-s)}\,,\ee
where $\tilde{\Lambda}^{2}$ can be quite large, since the integral
over the imaginary part of the polarization function is now
convergent. We may introduce $\tilde{J}_P(P^2,T)$ and
$\tilde{J}_V(P^2,T)$ as complex functions, since we now have both
the real and imaginary parts of these functions. We note that the
construction of either $\mbox{Re}\,J_P(P^2,T)$, or
$\mbox{Re}\,J_V(P^2,T)$, by means of a dispersion relation does
not require a subtraction. We use these functions to define the
complex functions $J_P(P^2,T)$ and $J_V(P^2,T)$.

In order to make use of Eq.\,(A16), we need to specify
$\tilde{J}_P(0)$ and $\tilde{J}_P(P_0^2)$. We found it useful to
take $P_0^2=-1.0$ \gev2 and to put $\tilde{J}_P(0)=J_P(0)$ and
$\tilde{J}_P(P_0^2)=J_P(P_0^2)$. The quantities $\tilde{J}_V(0)$
and $\tilde{J}_V(P_0^2)$ are determined in an analogous function.
This procedure in which we fix the behavior of a function such as
$\mbox{Re}\tilde{J}_V(P^2)$ or $\mbox{Re}\tilde{J}_V(P^2)$ is
quite analogous to the procedure used in Ref.\,[\,17\,]. In that
work we made use of dispersion relations to construct a continuous
vector-isovector current correlation function which had the
correct perturbative behavior for large $P^2\rightarrow-\infty$
and also described the low-energy resonance present in the
correlator due to the excitation of the $\rho$ meson. In
Ref.\,[\,17\,] the NJL model was shown to provide a quite
satisfactory description of the low-energy resonant behavior of
the vector-isovector correlation function.

We now consider the calculation of temperature-dependent hadronic
current correlation functions. The general form of the correlator
is a transform of a time-ordered product of currents, \be iC(P^2,
T)=\int d^4xe^{iP\cdot x}<\!\!<T(j(x)j(0))>\!\!>\,,\ee where the
double bracket is a reminder that we are considering the finite
temperature case.

For the study of pseudoscalar states, we may consider currents of
the form $j_{P,i}(x)=\tilde{q}(x)i\gamma_5\lambda^iq(x)$, where,
in the case of the $\pi$ mesons, $i=1,2$ and $3$. For the study of
scalar-isoscalar mesons, we introduce
$j_{S,i}(x)=\tilde{q}(x)\lambda^i q(x)$, where $i=0$ for the
flavor-singlet current and $i=8$ for the flavor-octet current
[\,18\,].

In the case of the pseudoscalar-isovector mesons, the correlator
may be expressed in terms of the basic vacuum polarization
function of the NJL model, $J_P(P^2, T)$ [\,18-20\,]. Thus, \be
C_P(P^2, T)=J_P(P^2, T)\frac{1}{1-G_{P}(T)J_P(P^2, T)}\,,\ee where
$G_P(T)$ is the coupling constant appropriate for our study of
$\pi$ mesons. We have found $G_P(T)=13.49$\gev{-2} by fitting the
pion mass in a calculation made at $T=0$, with $m_u = m_d =0.364$
GeV. The result given in Eq.\,(A18) is only expected to be useful
for small $P^2$, since the Gaussian regulator strongly modifies
the large $P^2$ behavior. Therefore, we suggest that the following
form is useful, if we are to consider the larger values of $P^2$.
\be \frac{C_{P}(P^2, T)}{P^2}=\left[\frac{\tilde{J}_P(P^2,
T)}{P^2}\right] \frac{1}{1-G_P(T)J_P(P^2, T)}\,.\ee (As usual, we
put $\vec{P}=0$.) This form has two important features. At large
$P_0^2$, ${\mbox{Im}\,C_{P}(P_0, T)}/{P_0^2}$ is a constant, since
${\mbox{Im}\,\tilde{J}_{P}(P_0^2, T)}$ is proportional to $P_0^2$.
Further, the denominator of Eq.\,(A19) goes to 1 for large
$P_0^2$. On the other hand, at small $P_0^2$, the denominator is
capable of describing resonant enhancement of the correlation
function. As we have seen, the results obtained when Eq.\,(A19) is
used appear quite satisfactory. (\,We may again refer to
Ref.\,[\,17\,], in which a similar approximation is described.)

For a study of the vector-isovector correlators, we introduce
conserved vector currents $j_{\mu,
i}(x)=\tilde{q}(x)\gamma_{\mu}\lambda_i q(x)$ with i=1, 2 and 3.
In this case we define \be J_V^{\mu\nu}(P^2,
T)=\left(g\,{}^{\mu\nu}-\frac{P\,{}^\mu
P\,{}^\nu}{P^2}\right)J_V(P^2, T)\ee and \be C_V^{\mu\nu}(P^2,
T)=\left(g\,{}^{\mu\nu}-\frac{P\,{}^\mu
P\,{}^\nu}{P^2}\right)C_V(P^2, T)\,,\ee taking into account the
fact that the current $j_{\mu,\,i}(x)$ is conserved. We may then
use the fact that \be J_V(P^2,T) =
\frac13g_{\mu\nu}J_V^{\mu\nu}(P^2,T)\ee and
\be\mbox{Im}\,J_V(P^2,T)&=&
\frac23\left[\frac{P_0^2+2m_u^2(T)}{8\pi}\right]
\left(1-\frac{4m_u^2(T)}{P_0^2}\right)^{1/2}e^{-\vec
k\,{}^2/\alpha^2}[\,1-2n_1(k)\,]\\
&\simeq& \frac{2}{3}\mbox{Im}J_P(P^2,T)\,.\ee (See Eq.\,(A7) for
the specification of $k=|\vec k|$.) We then have \be
C_V(P^2,T)=\tilde{J}_V(P^2,T)\frac1{1-G_V(T)J_V(P^2,T)}\,,\ee
where we have introduced \be\mbox{Im}\tilde{J}_V(P^2,T)&=&
\frac23\left[\frac{P_0^2+2m_u^2(T)}{8\pi}\right]
\left(1-\frac{4m_u^2(T)}{P_0^2}\right)^{1/2}[\,1-2n_1(k)\,]\\
&\simeq& \frac{2}{3}\mbox{Im}\tilde{J}_P(P^2,T)\,. \ee
In the literature, $\omega$ is used instead
of $P_0$ [\,4-6\,]. We may define the spectral functions \be\sigma_V(\omega,
T)=\frac{1}{\pi}\,\mbox{Im}\,C_V(\omega, T)\,,\ee and \be\sigma_P(\omega,
T)=\frac{1}{\pi}\,\mbox{Im}\,C_P(\omega, T)\,,\ee

Since different conventions are used in the literature [\,4-6\,],
we may use the notation $\overline{\sigma}_P(\omega, T)$ and
$\overline{\sigma}_V(\omega, T)$ for the spectral functions given
there. We have the following relations: \be
\overline{\sigma}_P(\omega, T)=\sigma_P(\omega, T)\,,\ee and
\be\frac{\overline{\sigma}_V(\omega,
T)}{2}=\frac{3}{4}\sigma_V(\omega, T)\,,\ee where the factor 3/4
arises because, in Refs. [\,4-6\,], there is a division by 4,
while we have divided by 3, as in Eq.\,(A22).

\section{}
\renewcommand{\theequation}{B\arabic{equation}}

Here we extend the work of Appendix A to consider case of finite
three-momentum, $\vec{P}$. We consider the calculation of
$\mbox{Im}J_P(P^0,\vec{P},T)$. The momenta $P^0$ and $\vec{P}$ are
the values external to the loop diagram. Internal to the diagram,
we have a quark of momentum $k+P/2$ leaving the left-hand vertex
and an antiquark of momentum $k-P/2$ entering the left-hand
vertex. It is useful to define \be
E_1(k)&=&\left|\vec{k}+\vec{P}/2\right|\\
&=&\left(k^2+\frac{P^2}4+kP\cos\theta\right)^{1/2}\ee and
\be E_2(k)&=&\left|\vec{k}-\vec{P}/2\right|\\
&=&\left(k^2+\frac{P^2}4-kP\cos\theta\right)^{1/2}\,.\ee Here
$k=|\vec{k}|$ and $P=|\vec{P}|$.

We have \be \mbox{Im}\,J_V(P^0,\vec{P},
T)=\frac12N_c\beta_V\,\epsilon(P^0)\mytint ke^{-\vec
k\,{}^2/\alpha^2}\left(\frac{2\pi}{2E_1(k)2E_2(k)}\right)\\\nonumber
\times\{[1-n_1(k)-n_2(k)]
\delta(P\,{}^0-E_1(k)-E_2(k))\\\nonumber-[n_1(k)-n_2(k)]
\delta(P\,{}^0+E_1(k)-E_2(k))\\\nonumber-[n_2(k)-n_1(k)]
\delta(P\,{}^0-E_1(k)+E_2(k))\\\nonumber-[1-n_1(k)-n_2(k)]
\delta(P\,{}^0+E_1(k)+E_2(k))\}\,.\ee Here, \be
n_1(k)=\frac1{e^{\,\beta E_1(k)}+1}\,,\ee and \be
n_2(k)=\frac1{e^{\,\beta E_2(k)}+1}\,.\ee In Eq. (B5), the second
and third terms cancel and the fourth term does not contribute. It
is useful to rewrite $\delta(P^0-E_1(k)-E_2(k))$ using \be
\delta[f(\cos\theta)]=\frac2{\left|\frac{\partial f}{\partial \cos
\theta}\right|_x}\delta(\cos\theta-x)\,,\ee where \be
x^2&=&\cos^2\theta\\&=&\frac{4P_0^2(k^2+P^2/4)-P_0^4}{4k^2P^2}\nonumber\,.\ee
We find \be \left|\frac{\partial f}{\partial \cos
\theta}\right|=\frac12kP\left|\frac{E_1(k)-E_2(k)}{E_1(k)E_2(k)}\right|\,,\ee
and obtain \be \mbox{Im}\,J_P(P^0,\vec{P},
T)=\frac12N_c\beta_P\,\epsilon(P^0)(2\pi)^2\int
\frac{k^2dk}{(2\pi)^3}e^{-k\,{}^2/\alpha^2}\\\nonumber
\int\frac1{2E_1(k)E_2(k)}[1-n_1(k)-n_2(k)]
\left|\frac{\partial{f(\cos\theta)}}{\partial{
\cos\theta}}\right|\\\nonumber
\times\delta(\cos\theta-x)d(\cos\theta)\,.\ee We note there is a
singularity when $E_1(k)=E_2(k)$. That occurs when $\cos\theta=0$
or $\theta=\pi/2$. For our calculations we eliminate the point
with $\theta=\pi/2$ when evaluating the angular integral over
$d(\cos\theta)\delta(\cos\theta-x)$ in the last expression. We
obtain \be \mbox{Im}\,J_P(P^0,\vec{P},
T)=N_c\beta_P\,\epsilon(P^0)\frac{4\pi^2}{(2\pi)^3}\int^{k_{max}}
k^2dk\,e^{-k\,{}^2/\alpha^2}\\\nonumber
\times\left.\frac{[1-n_1(k)-n_2(k)]}{kP|E_1(k)-E_2(k)|}\right|_x\,,\ee
where \emph{x} is obtained from Eq. (B9), \be
x=\frac{P^0}{kP}\left[k^2+\frac{P^2}4-\frac{P_0^2}4\right]^{1/2}\ee

\begin{acknowledgments}
We wish to thank Peter Petreczky for suggesting the work reported
here and for his explaining various features of the relevant
lattice calculations.
\end{acknowledgments}



\end{document}